# Stellar Double Coronagraph: a multistage coronagraphic platform at Palomar observatory

Michael Bottom,[1] J. Chris Shelton,[2] James K. Wallace,[2] Randall Bartos,[2] Jonas Kuhn,[2,3] Dimitri Mawet,[1,2] Bertrand Mennesson,[2] Rick Burruss, [2] Eugene Serabyn[2]

mbottom@caltech.edu

## ABSTRACT

We present a new instrument, the "Stellar Double Coronagraph" (SDC), a flexible coronagraphic platform. Designed for Palomar Observatory's 200" Hale telescope, its two focal and pupil planes allow for a number of different observing configurations, including multiple vortex coronagraphs in series for improved contrast at small angles. We describe the motivation, design, observing modes, wavefront control approaches, data reduction pipeline, and early science results. We also discuss future directions for the instrument.

*Subject headings:* High contrast imaging, Coronagraphy, Wavefront control

## 1. Introduction and Motivation

High contrast imaging is a rapidly evolving field, offering one of the most promising ways of obtaining spectra of extrasolar planets. Direct imaging of planets is challenging for two main reasons: first, stars are brighter than their orbiting planets by many orders of magnitude, and second, planets and their host stars are close to each other in angular separation, often only a few telescope resolution elements. In order to practically overcome these issues, at least three things are needed: an extreme adaptive optics system (ExAO) to reduce scattered starlight, a coronagraph to attenuate the diffraction pattern and allow observations very close to the host star, and post-processing image analysis to reduce quasi-static speckles due to remaining non-common path spatial and temporal optical imperfections. The large amount of scientific and technical effort towards improving these three areas has led to rapid developments in the field. While direct imaging has only discovered a handful of planets so far, there are recently commissioned instruments such as GPI (Macintosh et al. 2014), SPHERE (Zurlo et al. 2014), and SCEXAO (Jovanovic et al. 2015) that will expand the census of imaged planets around nearby stars.

---

[1]California Institute of Technology, MC 249-17, Pasadena, CA 91125, USA

[2]Jet Propulsion Laboratory, California Institute of Technology, Pasadena, CA, 91109 USA

[3](current address) Institute for Astronomy, ETH Zurich, Wolfgang-Pauli-Strasse 27, CH-8093 Zurich, Switzerland

When evaluating new coronagraphic systems and technologies, there are a few useful performance metrics to consider, including *contrast*, *inner working angle*, and *throughput*. For an unresolved, single star, contrast is the average brightness at a particular area of the image divided by the brightness of the star. Inner working angle is commonly defined as the angular separation where the flux of a putative companion is attenuated by the coronagraph to 50% of what it would be arbitrarily far away. The importance of inner working angle can be appreciated by noting that at a star 10 pc away, a planet at 1 AU will only be separated by 100 mas, just above one diffraction beamwidth for a 5 meter telescope operating at 2 $\mu$ m. Also, the areal discovery space around a star scales as the inverse square of the inner working angle, and the number of accessible stars in a survey complete down to an orbital radius goes as the inverse *cube* of the inner working angle (that is, halving the inner working angle allows one to observe a planet twice as far away at the same star-planet separation). Throughput refers to the fraction of planet light making it through the coronagraphic system as a function of angular separation. Throughput is also very important, as accumulating enough companion photons even at far separations from a star can be a challenge with faint planetary sources. Note that when comparing different coronagraph designs, many of these metrics are stated in units of diffraction beamwidths ($\lambda/D$) rather than absolute terms, since the merits of a design do not depend on the particular dish diameter.

One of the most promising technologies in the field is the vector vortex coronagraph (Mawet et al. 2005). Using a phase mask known as an optical vortex in an intermediate focal plane, for a clear aperture the coronagraph can provide very high contrast at an inner working angle of $0.9\lambda/D$, near the theoretical limit set by diffraction, with nearly 100% throughput at larger angles. Despite its advantages, a vector vortex suffers the same fate that all coronagraphs do when operating behind a telescope with a secondary mirror, sharply reduced contrast and degraded inner working angle due to diffraction from the secondary and any support struts in the telescope pupil.

There are certain ways to get around the limitations set by the secondary mirror and assorted support spiders. One way is to use a mask to re-image only a clear, unobscured pupil subaperture. This leads to severely reduced throughput and resolution; regardless, this method holds the record for closest directly imaged planet in units of diffraction beamwidths for a conventional coronagraph (Serabyn et al. 2010), though interferometric methods relying on non-redundant masking have done better (Kraus & Ireland 2012). Two other proposed designs also allow for improved contrast at low inner working angles when dealing with centrally obscured apertures. The first is based on a specially made pupil-plane apodizer upstream of the vortex (Mawet et al. 2013), which has the effect of redistributing the diffracted starlight in a way that it can be completely blocked in a post-coronagraphic Lyot plane, also at the cost of reduced throughput[1]; the second is by introducing another vortex in series with the first, which moves the light diffracted by the secondary to the center of the pupil, where it can be blocked (Mawet et al. 2011b). The SDC was built by the

---

[1] modern designs using pupil-remapping apodizers can recover most of this lost throughput, but are not used in this work

Jet Propulsion Lab[2] to allow for flexible development, testing, and on-sky evaluation and useful observing with these designs, as well as other ideas in wavefront control and coronagraphy, as will be elaborated below.

The layout of this paper is as follows. The next section of this paper discusses the physics and optics behind the vector vortex coronagraph and its various observing modes. Section 3 presents the optomechanical design of the instrument. Section 4 discusses the observing approach and data reduction pipelines for the instrument. Laboratory measurements and predicted performance are presented in Section 5. First on-sky results are presented in Section 6.

## 2. Background

The initial coronagraphic elements in the SDC are K-s band optical vortices, and most of the observing modes implemented in the instrument thus far involve their use in some way. In this section, we will summarize the principles behind optical vortices and their implementation as focal-plane coronagraphic elements. We also provide a theoretical description for the different instrumental observing modes. Readers may consult the cited papers below for more detailed discussions. We expect the SDC to evolve with time, adding functionalities and improving performance. Thus, what is presented here is a snapshot of the current state of the instrument.

### 2.1. Optical Vortices

An optical vortex is a device that creates an azimuthal phase ramp of $e^{il\theta}$, $l = 1, 2, 3...$ in light passing through it, where the integer $l$ is referred to as the "topological charge". A simple way to picture it is as a piece of glass whose thickness increases like a spiral staircase. As light passes through the spiral glass plate, it accumulates a different phase delay depending on its azimuthal position. For a particular choice of glass index and thickness profile, a smoothly increasing phase delay of 0 to $4\pi$ can be constructed; this is a "charge-2 vortex".

The spiral glass plate, also known as a "scalar vortex," is not the actual optic used in the SDC, as the machining tolerances are difficult to achieve, and the chromaticity of glass limits the usable bandwidth. There are three methods more commonly used to create high quality optical vortices, subwavelength gratings (Delacroix et al. 2014), photonic crystals (Murakami et al. 2013) and liquid crystal polymers. All these methods generate the phase delays with a spatially varying half-wave plate, the fast axis rotates twice as fast as the azimuthal coordinate. These are "vector vortices" since they use polarization to generate a phase delay through the Pancharatnam-Berry phase, rather than creating it through thickness variations in optics. The liquid crystal polymer

---

[2]Principal Investigator: Eugene Serabyn

method, the one we use, is discussed in Mawet et al. (2009).

## 2.2. Single vortex coronagraph

The principle behind the vortex coronagraph is the following. Light brought to a focus at the center of an optical vortex, experiences a total phase discontinuity, which creates a dark hole. If the vortex is placed at the focus of a telescope, the dark hole will propagate and expand at the center of the optical axis. The result is that at the following pupil, the dark hole will encompass the entire pupil–all the starlight will lie outside the aperture. A conventional pupil stop can then block this light. Any planetary companions that come to a focus *off* the center of the vortex will not experience the phase discontinuity, and will thus propagate almost normally. Figure 1 shows a schematic of this optical system.

It is useful to consider the vortex coronagraph in the optical Fraunhofer approximation, where there is a Fourier transform relationship between subsequent image and pupil planes. The input pupil is Fourier transformed to the image plane, then multiplied by the vortex function ($e^{il\theta}$, where $l$ is the topological charge), the *product* of which is Fourier transformed to the following pupil plane. In this view, the "perfect" performance of the unobscured vortex coronagraph is simply a statement that the Fourier transform of the product of an Airy disk and $e^{il\theta}$ (for even $l$) has no energy interior to the input pupil radius. In the case of a telescope with an on-axis secondary mirror (Mawet et al. 2011b), the electric field in the pupil following the vortex is

$$E_L(r) = \begin{cases} 0 & r < r_0 \\ -e^{i2\psi}\left(\frac{r_0}{r}\right)^2, & r_0 < r < R \\ e^{i2\psi}\left[\left(\frac{R}{r}\right)^2 - \left(\frac{r_0}{r}\right)^2\right] & r > R \end{cases}$$

where $r$ specifies the radial coordinate, $\psi$ specifies the angular coordinate, $R$ is the outer pupil radius, and $r_0$ is the radius of the secondary mirror. In the case of $r_0 = 0$, the unobscured case, the cancellation is perfect ($E_L = 0$) in the regime $0 < r < R$.

The perfect cancellation is only true for an Airy function input, not arbitrary circularly symmetric distributions of light. This is important because in cases of obscured apertures, the vortex will leak at a net level of $(r_0/R)^2$; see Figure 1. This may be improved somewhat using an oversized central mask in the Lyot plane, but at the cost of degraded throughput. It should be noted that the issue of pupil obscurations is a problem that inhibits the performance of all coronagraphic designs, and is not unique to the vortex coronagraph.

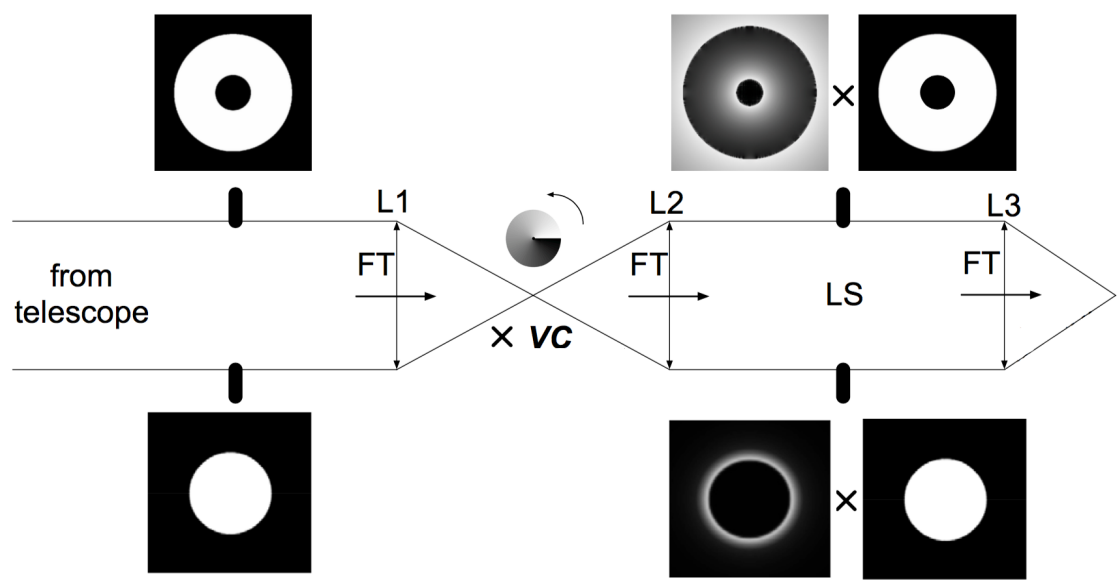

Fig. 1.—: The basic vortex coronagraph. Light from a clear telescope pupil (bottom row) is focused on the vortex by the telescope optics. At the following pupil, the starlight is moved to outside the pupil, where it is blocked by a Lyot stop. The dark pupil is then reimaged by the camera, with the starlight removed. For a centrally obscured aperture (top row), there is a residual halo of starlight left in the pupil.

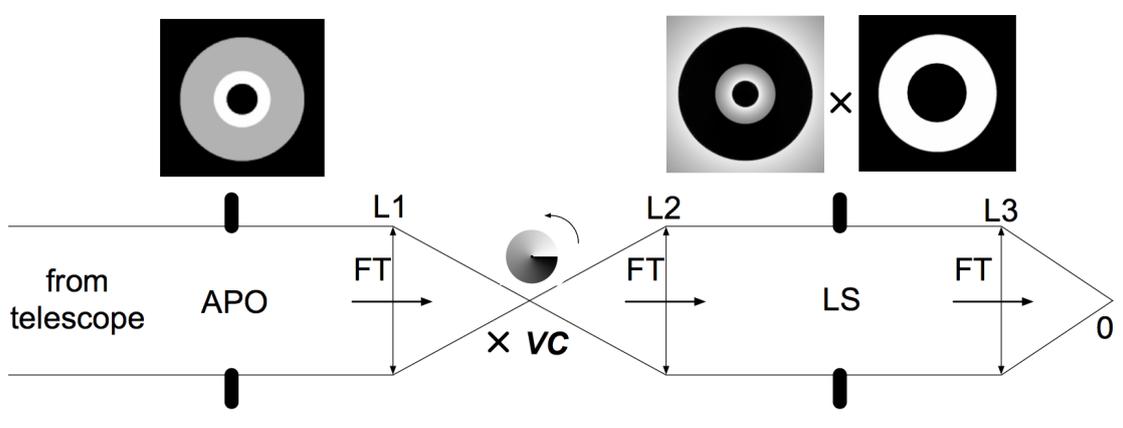

Fig. 2.—: The ring-apodized vortex coronagraph apodizes the input pupil of the telescope, with the result that in the pupil plane after the vortex, all the energy inside the pupil is localized to a ring (second square panel, top). A pupil stop can effectively block this light and thus have total starlight cancellation in principle.

### 2.3. Ring-apodized vortex coronagraph

One solution to the problem of obscured apertures can be provided by actually apodizing the *input* pupil with a partially transmissive ring (Mawet et al. 2013), see Figure 2. In the case of a transmissive ring apodizer, the electric field in the pupil following the vortex is

$$E_L(r) = \begin{cases} 0 & r < r_0 \\ -\left(\frac{r_0}{r}\right)^2, & r_0 < r < r_1 \\ (1-t)\left(\frac{r_1}{r}\right)^2 - \left(\frac{r_0}{r}\right)^2 & r_1 < r < R \end{cases}$$

where $r$ is the radial coordinate; $r_0$, $r_1$, and $R$ are the radius of the secondary, ring apodizer, and pupil, respectively; and $t$ is the transmissivity of the apodizer. By building the apodizer such that $t = 1 - (r_0/r_1)^2$, it is possible to create a region of total cancellation in the following pupil plane from $r_1$ to $R$. A Lyot stop can then remove the remaining light, interior to $r_1$. However, this comes at a cost of reduced throughput, as the size of the dark region is only a fraction of the pupil. For example, with a 35% central obscuration radius, the maximal throughput of the optical system can only be about 33%. Regardless, this approach can solve the problem of the secondary obscuration.

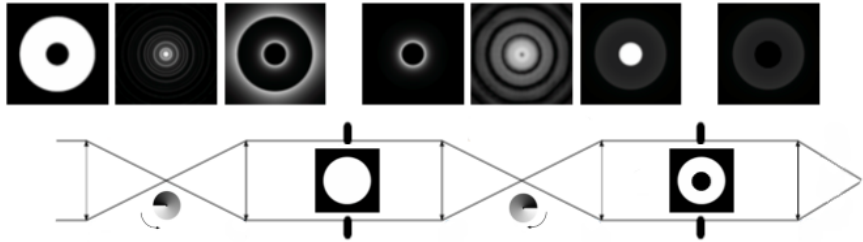

Fig. 3.—: Schematic of the dual-vortex coronagraph. The first vortex leaves a residual halo of light (4th panel from left) which is moved behind the pupil by the second vortex (2nd panel from right).

## 2.4. Multistage vortex coronagraph

Another approach to mitigating the problem of secondary obscurations is by having a second vortex in series with the first (Mawet et al. 2011b). The residual starlight from the first vortex is "folded back" by the second vortex to behind the secondary obscuration in the pupil plane, where it can then be blocked. In this case, the electric field after the final Lyot stop is

$$E_L(r) = \begin{cases} 0 & r < r_0 \\ \left(\frac{r_0}{R}\right)^2, & r_0 < r < R \\ 0 & r > R \end{cases}$$

where $r$ is the radial coordinate, $r_0$ is the radius of the secondary obscuration, and $R$ is the radius of the pupil. The electric field is not completely nulled, but reduced in amplitude significantly, with the leakage being reduced to $(r_0/R)^4$. In addition to the imperfect cancellation, another disadvantage of this method is the precise alignment needed between the two vortices. The significant advantage of the multistage coronagraph, however, is the high throughput.

## 3. Design of the Stellar Double Coronagraph

The SDC was built to demonstrate and use these and other coronagraphic approaches on-sky. It operates between the P3K adaptive optics system and the near-IR imager PHARO on the Hale 200" telescope; see Figure 4. The dimensions of the instrument are 25.4cm x 45.7cm x 91.4cm, and the weight is about 70 kg.

The SDC accepts the f/15.7 output beam from the adaptive optics system. An input fold mirror (Figure 5, top right) at a compound angle steers the beam towards the coronagraph. Before the coronagraph, an infrared dichroic splits off J-band light towards a quad cell tracker (see next paragraph). After passing through the first vortex phase mask, the light hits an off-axis paraboloid (Figure 5, top left), where it is collimated and then bent towards the first Lyot stop at the first pupil plane (green rectangular base). The optics in the Lyot plane are designed to be mounted at a 5 degree angle, so that the reflected (rejected) light returns at a different angle to the previous fold mirror. An internal camera with a flip lens (Figure 5, bottom left) uses this broadband infrared light to image the pupil (or object), assisting with initial alignment. The light passing through the Lyot stop is folded to a second off-axis paraboloid, which focuses it onto the second vortex phase mask, mounted on an identical linear slide. The second off-axis paraboloid controls the focus and vertical image position on the second vortex mask, and the horizontal position is controlled by the slide itself–this avoids changing the off-axis angle of the paraboloid. After the second vortex, another off-axis paraboloid collimates the beam to a second Lyot plane, where a second (reflective) Lyot stop may be installed. The collimated beam hits a final off-axis paraboloid (Figure 5, top right), and the converging beam is bent towards the infrared imager PHARO by an output fold mirror. A summary of the degrees of freedom and controls is presented in Table 2 of the Appendix.

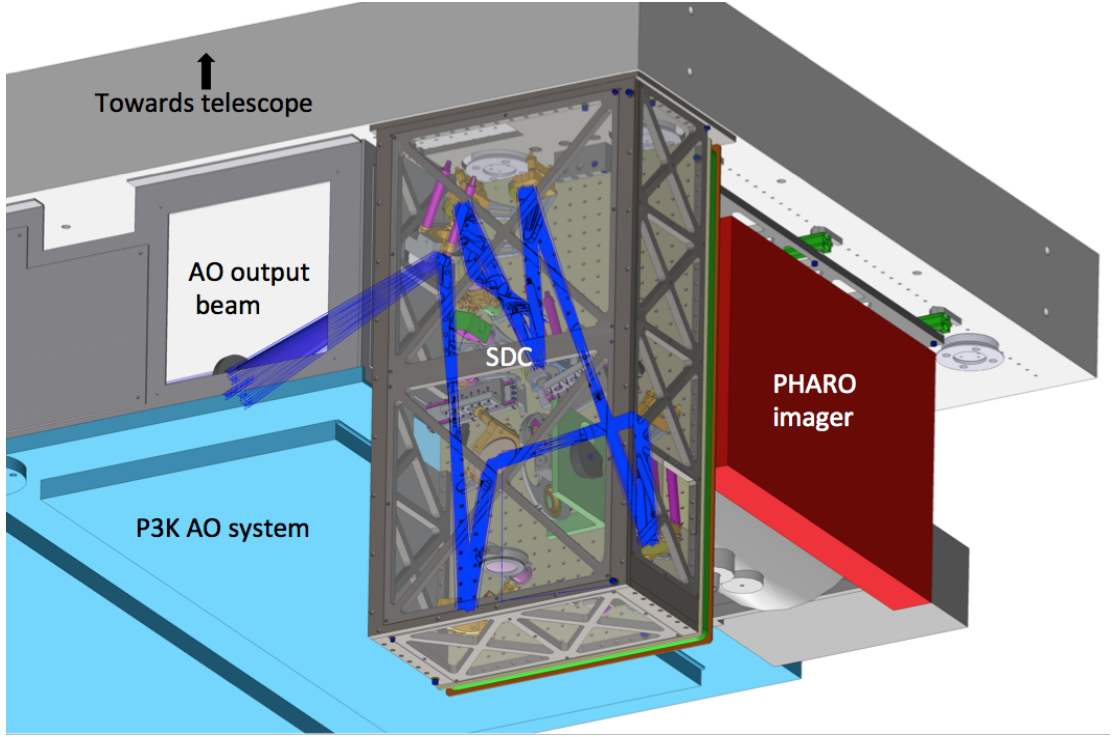

Fig. 4.—: The SDC mounts between the P3K adaptive optics system (blue rectangle, left) and infrared imager PHARO (partially visible behind its red electronics box). The imager, coronagraph and adaptive optics system all attach to Cassegrain port of the telescope.

A very important source of error in a coronagraphic system is tip/tilt "leakage" error ie,

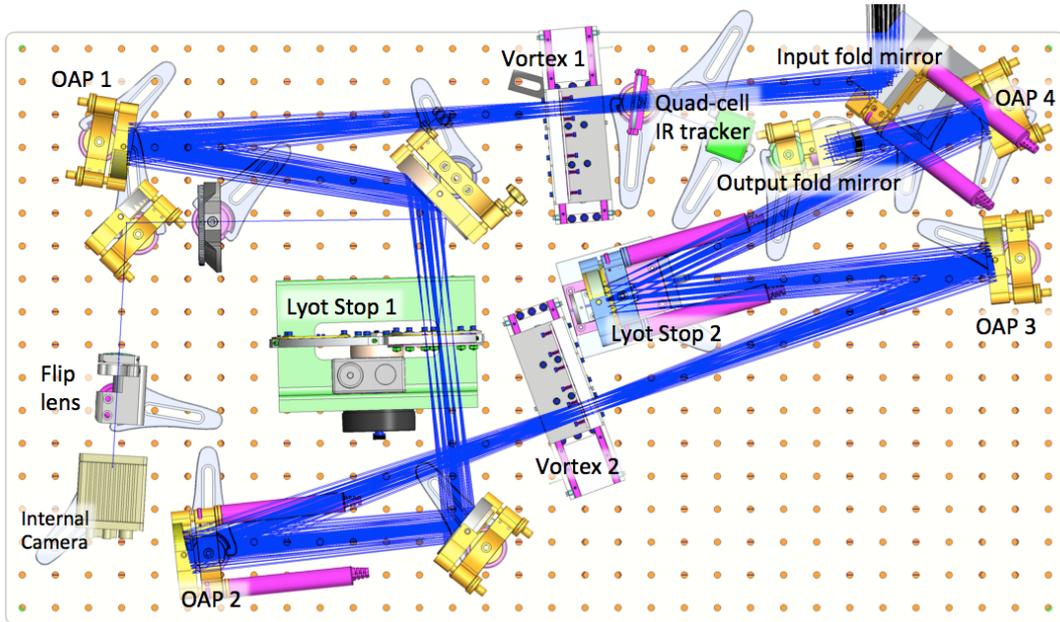

Fig. 5.—: The optomechanical layout of the SDC; refer to the text for a more detailed description. Following the input beam from the top right of the figure: first fold mirror, dichroic beam splitter, linear coronagraphic slide, off-axis paraboloid, fold mirror, Lyot plane, fold mirror, off-axis paraboloid, linear coronagraphic slide, off-axis paraboloid, Lyot plane, off-axis paraboloid, fold mirror. The infrared tracker is the green square. The image and pupil viewing camera and lenses are shown on the left, directly below the first off-axis paraboloid. In this orientation, the output beam to the infrared imager PHARO exits downward into the page.

misalignment of the input image with the coronagraph. In a Cassegrain instrument, this is especially challenging as the differential flexure between the coronagraph and wavefront sensor of the adaptive optics system means that the star will slowly drift off the coronagraph through the course of an observation. Our solution to this problem is a custom quad-cell infrared tracker. A dichroic splitter sends J-band (1.1-1.4 $\mu$m) light to the infrared tracker (green box at top of Figure 5), and lets the science wavelengths ($H$ and $K$ band) pass to the coronagraph. The tracker is positioned to be very close to the coronagraphic optical element, so differential flexure is minimal. The output of the tracker is digitized, analyzed, and locked to the adaptive optics system in a proportional-integral-derivative type controller, with a variable update time. Typically, corrections are sent once every 5-20 seconds, depending on observing conditions, as the goal is to correct slow non-common path drifts. Using a Hamamatsu G6489-01 quad cell as the sensor, the tracker delivers a positional accuracy of better than 3 mas, and has a linear range of $\pm$ 250 mas. The tracker uses a variation of the denominator-free centroid technique (Shelton 1997), allowing acquisition of targets well outside this linear range or the AO field stop size. Its dynamic range corresponds to stars between 0 and about 8.5 magnitudes, the fainter magnitude approximately coinciding with the high-Strehl ($>$80%) cutoff of the P3K adaptive optics system.

There are a few design choices that make the instrument quite flexible for adapting new configurations. The mounts holding the coronagraphs are connected to the motorized linear stages by magnets, which can be quickly removed and replaced, with a positional repeatability of a few microns. Furthermore, each mount accommodates up to three separate coronagraphic elements, which can be slid in and out remotely, although one slot is typically occupied by a single-mode alignment fiber. The Lyot wheel has five slots, and the face of the wheel can be quickly removed for installation of new pupil masks. Finally, the second Lyot plane can also accept pupil masks, which can be positioned accurately via the use of magnets installed on the optical mount. Some of these optics are shown in Figure 5.

## 4. Observing sequence and data reduction

### 4.1. On-sky calibrations and observing strategy

The Hale telescope has an equatorial mount, so Cassegrain instrument gravity vectors vary with time and depending on the target. In practice, this changing vector leads to two important effects, which are linked. First, the AO system output Strehl ratio degrades after a large slew; second, the speckle pattern changes due to flexure. These two effects can be viewed as introduction of low-order and high-order aberrations, respectively. We compensate for these two effects separately in two calibration steps that we perform after executing a large slew of the telescope. These steps are performed using the internal white light source of the adaptive optics system.

### 4.1.1. Correction of non-common path low-order aberrations: MGS

The Strehl ratio at the first coronagraphic mask is important, as the total performance of that vortex depends sensitively on the amount of low-order aberration present. We compensate for the Strehl degradation with telescope pointing by performing the Modified Gertzberg-Saxton (MGS) phase retrieval algorithm (Burruss et al. 2010) in the detector plane after slewing to a new target, using the internal AO white light source. This technique calculates phase errors using a set of defocused PSF, iteratively updating the shape of the deformable mirror to maximize the Strehl ratio. While the Strehl ratio at the detector is not the same as on the first vortex mask focal plane, the SDC internal Strehl ratio, i.e., that measured from the first coronagraphic focal plane to the detector, is typically about 95%. Therefore, it is safe to state that the MGS algorithm's optimization of low-order aberration between the deformable mirror and the detector focal plane achieves a high Strehl PSF on the *coronagraphic mask* plane as well. Using MGS, $K_s$ Strehl ratios are typically improved from about 70% to 95% from the input of the adaptive optics system to the detector after large slews or other disruptive events.

### 4.1.2. Correction of non-common path high-order aberrations: speckle nulling

The most important factor limiting contrast in high contrast imaging systems are speckles, quasi-static bright spots in the focal plane caused by non-common-path phase and amplitude aberrations in or after the beamsplitter that separates science light from light to the AO wavefront sensor. The total intensity of these speckles (that is, the speckle flux in the focal plane from 0-$33\lambda/D$) is very low in absolute terms, as they correspond to wavefront errors with amplitudes of 5-25 nm, but they are typically much brighter than any planetary companions, at raw contrasts of $10^{-3}$ - $10^{-4}$ in $K_s$. They are also sensitive to gravitational flexure in the optical train that evolves on ∼minute timescales. The MGS algorithm has difficulty correcting these low intensity, high spatial frequency speckle aberrations, as the signals-to-noise of these errors are very low in the out-of-focus images that the procedure uses to determine wavefront quality. Despite the fact that the detector readout time is a few seconds for PHARO, these slowly-evolving static speckles may be reasonably tackled using the science camera as a sensor.

To lower speckle intensity, we have implemented a speckle nulling code (Savransky et al. 2012). The technique involves using the deformable mirror to generate spots at the exact locations of bright speckles in the image, then varying the phase of the electric field of the spots. By measuring the modulation in the intensity due to the interference of the spot and speckle, it is possible to calculate the phase of the speckle, and hence cancel it with the deformable mirror.

A typical iteration proceeds as follows. First, speckles are identified in the image using a local maximum filter, and aperture photometry is used to get their intensities. The position on the detector and intensity of an individual speckle is converted to an equivalent sinusoidal shape on the deformable mirror, with the spatial frequency and orientation determining the position, and the amplitude determining the brightness. The phase of the sinusoid on the deformable mirror is changed four times, with the intensity of the speckle measured at each step. These intensities are then used to determine the phase that would cancel the speckle completely, the "null phase." In order to compensate for deformable mirror hysteresis, intensity calibration imperfections, and other possible errors, a second loop then modulates the amplitude of each null sinusoid in four steps (0%, 33%, 66%, 110%), with the speckle intensity recorded once again and used to determine the optimal amplitude of the sinusoid on the deformable mirror. While the second loop is not strictly necessary, and is not used for the first few iterations, it is very useful as it prevents speckle amplification if the phase is incorrectly measured. It also helps define a stop condition–when the calculated optimal gains start dropping to zero, there is no benefit to continue running the loop. All in all, either four or eight images are required per iteration.

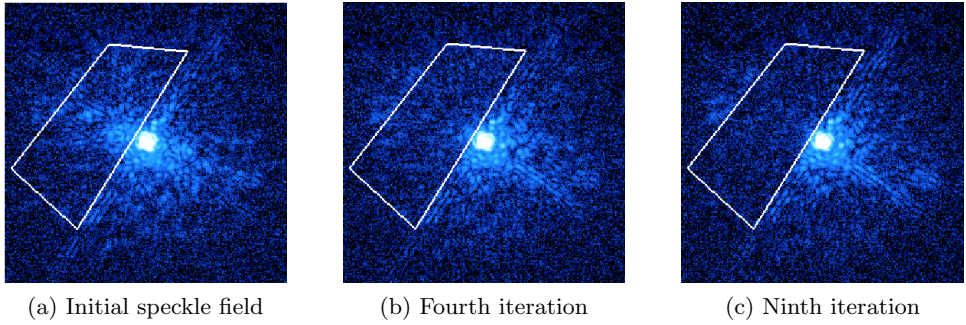

(a) Initial speckle field  (b) Fourth iteration  (c) Ninth iteration

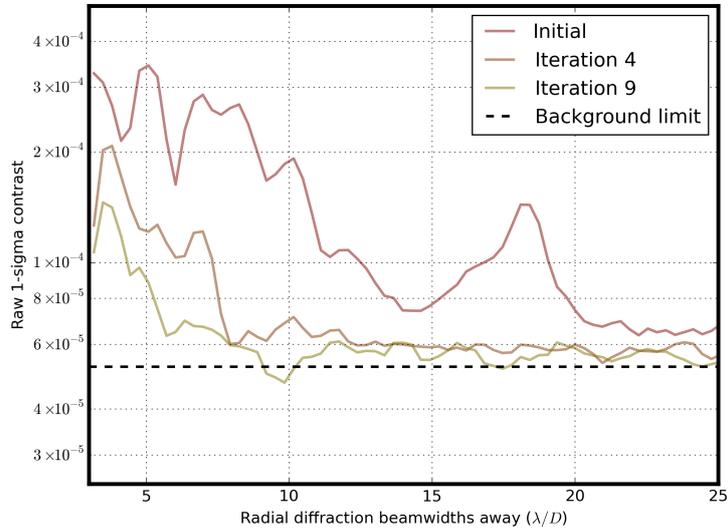

(d) Contrast in region

Fig. 6.—: Coronagraphic speckle nulling in the dual-vortex mode using the internal white light source of the adaptive optics system; see 4.1.2. (a) The initial results of PSF correction using MGS (4.1.1) still leaves many residual speckles in the focal plane. (b) Four iterations of speckle nulling remove most of the residual wavefront errors (c) Nine iterations get to within a factor of two of the detector read noise from 5 - 25 $\lambda/D$. The white polygon demarcates the control region, which is selected by mouse clicks in the half-region control mode. (d) Contrast improvement measured in the control region shows factors of 3-6 improvement, which are significant for companion detectability. The contrast curve is defined in the usual way, with the standard deviation (ie, 1 $\sigma$) of surface brightness at each radial separation being used to generate the curve, and normalized by dividing by the peak flux of the non-coronagraphic PSF (not shown). The background limit is determined by the contrast in a region of the detector 100's of $\lambda/D$ away. The preprocessing steps performed on the data only consist of dark subtraction and flat-fielding.

After a few iterations, the internal instrumental contrast through the adaptive optics system improves typically by a factor of 3-6 when nulling half of the focal plane, and less when nulling a centro-symmetric region. The reason for the difference is that in the half aperture case, both phase and amplitude errors can be corrected with the single deformable mirror; in the full aperture case, only phase variations can be corrected. The system-level description of the speckle nulling code is shown in Figure 7.

### 4.2. Observing strategy

Similar to other high contrast imaging programs, our observing strategy is driven by the need to reduce residual speckles. There are many different methods to deal with speckles, including angular differential imaging (Marois et al. 2006a) and spectral differential imaging (Marois et al. 2006b). The former is precluded by the equatorial/Cassegrain mount configuration, and the latter because our imager is not an integral field spectrograph.

The strategy we use is the classical approach called "reference differential imaging" (Mawet et al. 2011a), and involves quickly alternating observations between our target star and a reference star. The reference star is selected to be nearby, and to have a similar visible magnitude and spectral type. The proximity of the star ensures that the gravity vector is similar, leading to a nearly identical speckle pattern. The visible magnitude and spectral type ensure that the AO correction and brightness in the detector plane are very similar. Dithering between the target and reference on a few-minute cadence means that the slow temporal evolution of the speckle pattern is thus tracked. With datacubes of the target and reference, it is possible to remove speckles in post-processing, as will be described below.

### 4.3. Data reduction pipeline

Data reduction is an integral part of high contrast imaging, and significant advances in performance have come from new algorithms and reduction methods (Lafrenière et al. 2007; Soummer et al. 2012). Understanding of reduction code behavior and performance have a direct bearing on observing strategies.

The SDC pipeline is an automated high contrast imaging reduction code written in Python. It performs the low-level preprocessing of the images, registration, and various forms of point spread function subtraction, as will be explained below.

The pipeline begins by building a database of each observing sequence from logsheet entries, taking as input exposure numbers. It checks for errors in the logsheet such as mislabeled filters, stops, etc, by examining file headers. A bad pixel mask is generated from flat fields of varying fluxes, bad pixels are treated with a spatial median filter. Dark subtraction/flatfield normalization is performed after this point. Image registration is done to the subpixel level by a discrete Fourier-transform upsampling algorithm described in Guizar-Sicairos et al. (2008), with shifting done using linear interpolation. During observing, we generate astrometric reference spots by adding a checkerboard pattern to the deformable mirror, corresponding to the outer spatial frequency controlled by the mirror. In practice, this checkerboard is generated on-sky by changing the convergence points of the wavefront sensor, with an amplitude corresponding to 5% of the stroke of the deformable mirror.

Using the registered reference and target image cubes, three independent point-spread function subtractions are performed. The first is classic PSF subtraction, where an intensity-scaled median reference PSF is subtracted from the median target PSF. The second method is a full-frame principal components analysis, also known as the Karhunen-Loeve eigenimage decomposition (KLIP) algorithm, described in Soummer et al. (2012). Here, each target image is projected onto a low-dimensional subspace derived from the principal components of the *reference* image library. The low-dimensional projection is subtracted from the input data. If an off-axis source is present in the target, but not the reference, it should largely be unaffected after subtraction of the principal components, though modeling of the algorithm's effects is necessary. One issue with using full frame KLIP is that the noise properties vary significantly as a function of radius from the star, and principal components analysis does not work well with this spatially varying noise. The way we compensate for this is to partition the image into small, overlapping zones and perform KLIP on each zone, then take the median to reconstruct the image. The zones are chosen to be small enough such that the noise is fairly uniform over each of them.

In addition to the reference cube, a "supercube" is constructed by choosing the most highly similar images to the target datacube over the entire observing run (not just the target and reference frames), selected via image correlation coefficients. Using the supercube as the reference, the data is reduced again; this typically gives a 30-40% improvement in contrast at close inner working angles. Unsurprisingly, most of the frames selected by the supercube generating algorithm are from the calibrator star.

| Observing mode | 1st coronagraph | First Lyot | 2nd Coronagraph | Second Lyot |
|---|---|---|---|---|
| Open | Open | Open | Open | None |
| Single Vortex | Ks band vortex | Lyot stop 1 | Open | PHARO pupil stop |
| Ring-apodized vortex | Open | Ring apodizer | Ks band vortex | Lyot stop 2 |
| Dual Vortex | Ks band vortex | Lyot stop 1 | Ks band vortex | PHARO pupil stop |

Table 1:: The different on-sky observing modes of SDC; see section 2 for a description of each of these modes.

## 5. Coronagraph configurations and laboratory performance

The SDC currently has four operational on-sky observing modes, listed below in Table 1. The "open" observing mode is used for flat-fielding exposures and non-coronagraphic observing. The "single-vortex" mode is used for comparing contrast ratios between the two vortices, but is not generally used for science as it offers no observational advantage due to the effect of the secondary. The "ring-apodized mode" and "dual vortex mode" are used for science observations, with the majority of time thus far going to the dual vortex mode. The reasons for this are twofold. First, the sensitivity to tip/tilt errors is much higher in the ring-apodized mode compared to dual-vortex mode ($\theta^2$ vs. $\theta^4$). This problem is compounded by the fact that the tip/tilt pointing control is not as good, as the tracker is located further away from the vortex mask used (that being in the *second* SDC focal plane, after the first Lyot wheel housing the apodizer). Furthermore, despite having theoretically perfect rejection, the theoretically maximal throughput of the ring-apodized coronagraph is only about 33%, as mentioned earlier.

We first measured performance of the observing modes in controlled laboratory conditions. In this case, the same P3K adaptive optics system, coronagraph, and detector were used, but the light source was the internal single-mode fiber white light source from the AO system. Additionally, all the tests were performed at a stationary, vertical gravity vector.

### 5.1. Single and dual vortex observing modes

The results comparing single and dual vortex modes are shown in Figure 8. The dual vortex mode provides a dramatic improvement in inner working angle, as shown by the large boost in contrast from 1 - 2 $\lambda/D$. Contrast is also improved at 3 - 10 $\lambda/D$, with the diffraction rings substantially removed. The point-spread functions are shown at the top; note the dual vortex PSF closely reproduces the original Airy function PSF, at a reduced intensity, as expected from Figure 3, where the output pupil is a copy of the input pupil but fainter. The measured peak rejection of about 100:1 is consistent with theoretical prediction of 80:1 for the Palomar secondary/primary mirror size ratio; the "better" than expected performance is mainly due to the fact that there is a small 25 $\mu$ chromium dot at the center of the vortex to compensate for manufacturing imperfec-tions that can lead to stellar leakage. We note that radial contrast, not peak rejection, is a true measure of coronagraph performance, but peak rejection provides a quick way to check whether the coronagraph is working to design expectations.

## 5.2. Ring-apodized vortex observing mode

The ring-apodizer was installed and tested on-sky in February 2015. Lab tests indicated performance consistent with theoretical expectations (see Figure 9 (b)). The advantages are high starlight suppression at small inner working angles. (For a telescope design with a 20% secondary obscuration, such as Keck, the throughput would be about twice as high, making this approach much more advantageous). Regardless, the lab tests demonstrate the validity of the concept, with a close match between theory and measurement. However, the minimum rejection is about 1000:1 at the peak of the psf, as opposed to infinite. There are a few explanations for the imperfect rejection, such as imperfect Strehl at the position of the second vortex. Additionally, the central chromium spot on the vortex is not taken into account in the idealized performance model. Similarly, bright light from defective pinned actuators in the AO system were clearly visible in the "null" region of the mask. Finally, interferometric testing revealed a 20 nm phase difference between the opaque and transmissive annuli in the apodizer, though at the operational wavelength of over 2 $\mu$m, this is unlikely to be significant.

## 6. On-sky performance

First light observations with SDC took place in February-March 2014, with full science observations in dual-vortex mode commencing in October 2014. Other observing runs were in Feb 2015 (nearly all lost to weather) and May 2015. This section will present some of the early engineering and science results of the instrument, some of which have been already published. It is not the intent of this paper to present the complete analyses of all our science targets, but a few preliminary results showcasing the performance of the instrument are summarized.

In general, sky performance (as measured by expected peak-to-peak rejection ratio) is within a factor of two of the lab-measured values. Much of the deviations from theory can be attributed to the imperfections in adaptive optics performance, such as imperfect Strehl ratio, bad actuators, etc. Internal rigidity of the instrument is generally very good, with the two focal-plane vortex masks staying co-aligned throughout observations despite large slews between different target stars. An example of a measurement of one of our survey targets is shown in Figure 10, with the image showing low residual diffraction with no evidence of companions.

### 6.1. Confirmation of physical association of epsilon Cephei b

Epsilon Cephei (HD 21136) had a previously reported (stellar) companion about 50 times fainter at 330 mas separation (Mawet et al. 2011a), but it was unknown whether it was physically associated or a background alignment. The close separation measured in 2010 (3.6 $\lambda/D$ for a 5 meter aperture) made it an attractive target for our first science observations. Additionally, it would allow more accurate measurement of position as the angular resolution of the full 5 meter dish would be used; the previous observation was performed in 2010 with a 1.5 m clear sub-aperture of the Hale telescope. We observed the star on 13 October 2014, and were able to see the companion in the raw coronagraphic image, with no post-processing (Figure 11). Classical PSF subtraction was all that was needed to measure astrometry and photometry. Given that the proper motion of Epsilon Cephei is more than 400 mas/yr, and our measured companion separation is $216 \pm 6$ mas (2.1 $\lambda/D$) after more than four years, physical association is definitively confirmed. Significant orbital motion is also evident, as the position angle has changed from $90 \pm 10$ degrees to $66 \pm 3$ degrees, and the orbit separation decreased from $330 \pm 50$ mas to the currently measured value. The data confirms that the orbit is clearly far from edge-on, but the two data points do not allow for a detailed characterization.

### 6.2. Identification of the "compact object" companion to delta Andromeda

On the initial science run, we observed the star delta Andromeda, which is a spectroscopic binary with a period of about 58 years. The companion had been previously hypothesized to be a white dwarf (Gontcharov & Kiyaeva 2002; Judge et al. 1987), but had never been imaged due to its faintness and proximity to the primary. Thanks to the high contrast at low inner working angles, we were able to easily detect the companion in the raw image. The companion separation was found to be about 360 mas (4 $\lambda/D$) with a contrast of 6.2 magnitudes in K-band. Bottom et al. 2015 showed that the companion was much too bright to be a white dwarf, and was more likely a main sequence star of K-type. Again, the advantage of high contrast at low inner working angles allowed for a robust detection and characterization; see Figure 12.

## 7. Conclusion and future work

We have presented the motivation, design, and current performance of a new multistage coronagraphic instrument at Palomar observatory's 200" Hale telescope. The SDC is currently the only multistage coronagraph in operation, and has also successfully tested the ring-apodized vortex coronagraph concept, a promising way of pursuing high contrast at low inner working angles when behind obscured telescope pupils. The SDC is fully operational and actively pursuing both astronomical observations and new technical developments at the same time.

Other observational and wavefront sensing modes can be envisioned, and will be implemented as funding and teaming arrangements allow. Potential coronagraphic modes include other focal plane mask-based coronagraphs, such as the band-limited Lyot coronagraph (Kuchner & Traub 2002), and also pupil-plane coronagraphs such as the shaped pupil (Kasdin et al. 2003) and phase-apodized coronagraph (Snik et al. 2012). Shorter-wavelength operation is also possible, as is a nulling interferometry mode. In the wavefront sensing area several steps are conceivable, including phase-shifting interferometry for direct measurement of speckle phases (Serabyn et al. 2011), Lyot-plane wavefront sensing (Singh et al. 2014), and speckle phase measurements with the self-coherent camera approach (Galicher et al. 2008). Finally, post-coronagraphic spectroscopy is also now enabled at Palomar, first with PHARO itself using the internal grisms, and potentially with other spectrometers, such as upcoming energy-resolving MKID detectors (Mazin et al. 2014).

## 8. Acknowledgements

We are pleased to acknowledge the Palomar Observatory staff for their enthusiastic and excellent support. We thank the referee for a careful and thorough read, and comments which improved the paper. MB is supported by a NASA Space Technology Research Fellowship, grant NNX13AN42H. Part of this work was carried out at the Jet Propulsion Laboratory, California Institute of Technology, under contract with the National Aeronautics and Space Administration (NASA).

## 9. Appendix

| Element | Degrees of Freedom | Controls |
| --- | --- | --- |
| AO output beam | Tip, Tilt | X, Y on focal plane masks |
| Input fold mirror | X, X+Y | Lateral input pupil position |
| Focal plane mask 1 slide | X | Mask 1 X position |
| First Lyot Wheel | $\Theta$ | Lyot stop choice |
| OAP #2 | X, Y, X+Y | Mask 2 Image Y position/focus |
| Focal plane mask 2 slide | X | Mask 2 X position |
| Lyot plane #2 | X, Y, X+Y | Image position on detector |
| Output fold mirror | X, Y | Output pupil position |
| Flip lens | In/Out | Image/Pupil on internal camera |

Table 2:: The actuated optics in the SDC, their degrees of freedom, and the optical fields they control. Refer to Figure 5 for the an optical layout

- 19 -– 19 –

| Item | Vendor | Part | Notes |
| --- | --- | --- | --- |
| Vortices | JDSU | | 25 $\mu$m dot in center |
| Optics mounts | Newport/Thorlabs | Assorted 2 and 3" | |
| Actuators | Newport | TRA6 | 0.2 $\mu$m step, 1.5 $\mu$m abs. |
| Actuator controller | Galil Motion Control | DMC-4183 | |
| Lyot wheel | | Custom | 5 slots |
| Lyot controller | Sigma-Koki Co. | PAT-001 | 250,000 cts/rev |
| Vortex stage controllers | Applied Motion Products | ST5 | Had to filter PWM signal |
| Internal camera | Sensors Unlimited | InGaAs | $J$, $H$, $K$ sensitivity |
| Piezo stage | Physik Instruemnte | P-752 | 0.2 nm resolution |
| Piezo controller | Physik Instrumente | E-516 | Few nm resolution |

Table 3:: List of optics, electronics, and related information

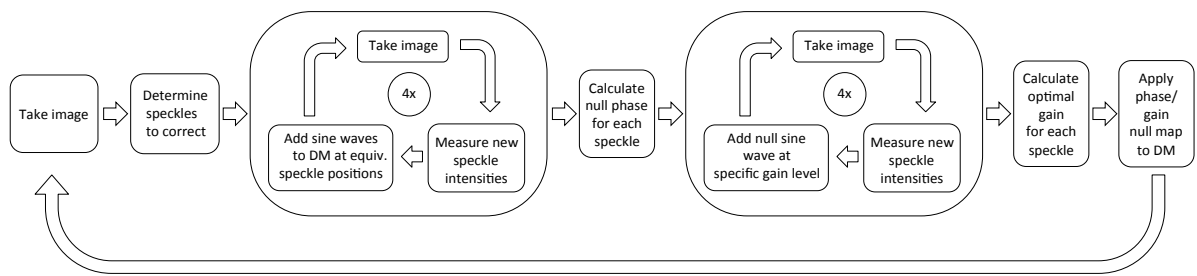

Fig. 7.—: Outline of the speckle nulling loop. See Section 4.1.2 for a description of the loop.

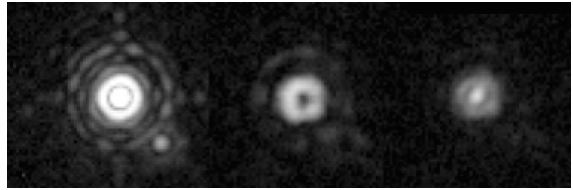

(a) Zero, single, and dual vortex PSFs; same (linear) scale

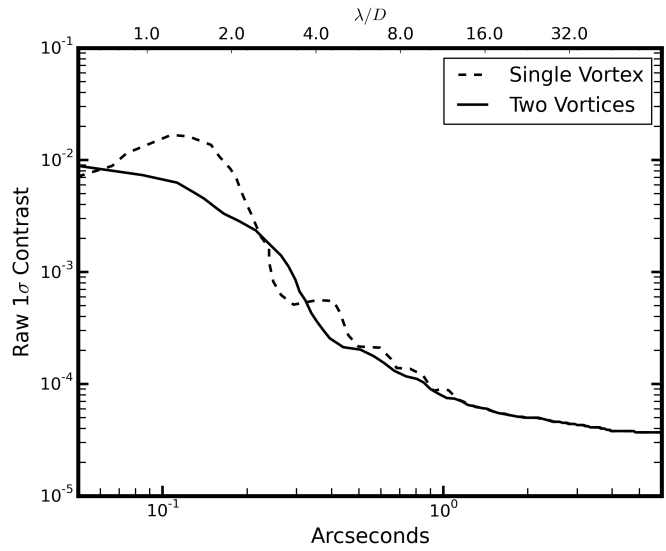

(b) Raw contrast

Fig. 8.—: Laboratory contrast measurement comparing single (dashed curve) and dual vortex mode (solid curve). The contrast curve is defined in the usual way, with the standard deviation (ie, $1\,\sigma$) of surface brightness at each radial separation being used to generate the curve, and normalized by dividing by the peak flux of the non-coronagraphic PSF (not shown). The preprocessing steps performed on the data only consist of dark subtraction and flat-fielding.

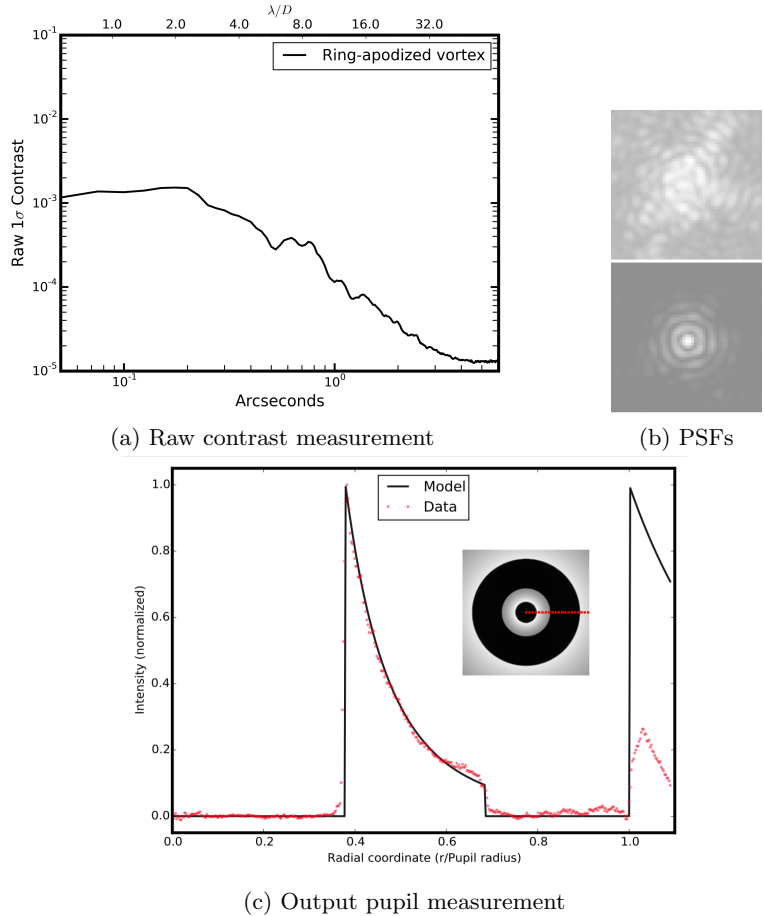

(a) Raw contrast measurement  (b) PSFs

(c) Output pupil measurement

Fig. 9.—: (a) Raw contrast measurement with the ring-apodized vortex coronagraph. The contrast curve is defined in the usual way, with the standard deviation (ie, 1 $\sigma$) of surface brightness at each radial separation being used to generate the curve, and normalized by dividing by the peak flux of the non-coronagraphic PSF (not shown). The preprocessing steps performed on the data only consist of dark subtraction and flat-fielding. (b) The coronagraphic (top) and non-coronagraphic PSF, shown on different logarithmic scales to enhance features. (c)The measurement of the output pupil intensity corresponds well to theoretical expectations, with the major discrepancy being outside the pupil. This is due to the presence of an chromium dot in the center of the vortex, reducing stellar leakage. The center of the PSF is the brightest, so light blocked there will not show up outside the pupil.

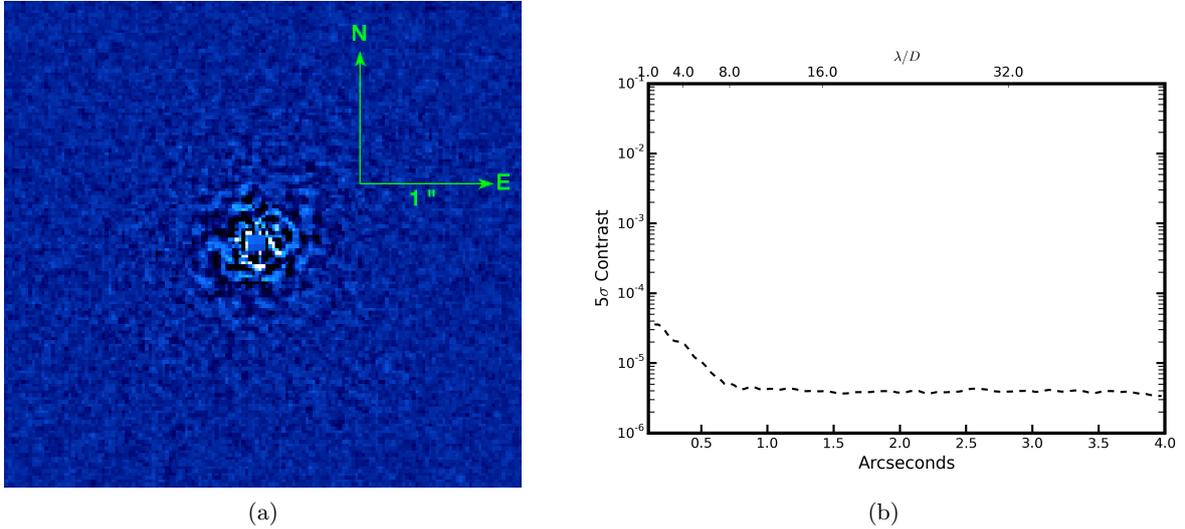

(a)                  (b)

Fig. 10.—: A reduced image of one of our target stars ($K=8$, $V=6$) with the associated $5\sigma$ contrast curve on the right. The reduction strategy used was a zonal principal components analysis (KLIP) algorithm (see Section 4.3), with the principal components generated from a calibrator star with similar brightness and $V-K$ color. The total open shutter time on this target was 14 minutes, with the same time on the calibrator star (backgrounds, flats, and non-coronagraphic PSF frames were recorded separately). This measurement did not involve speckle nulling, so contrast at small angles can likely be improved further in the future.

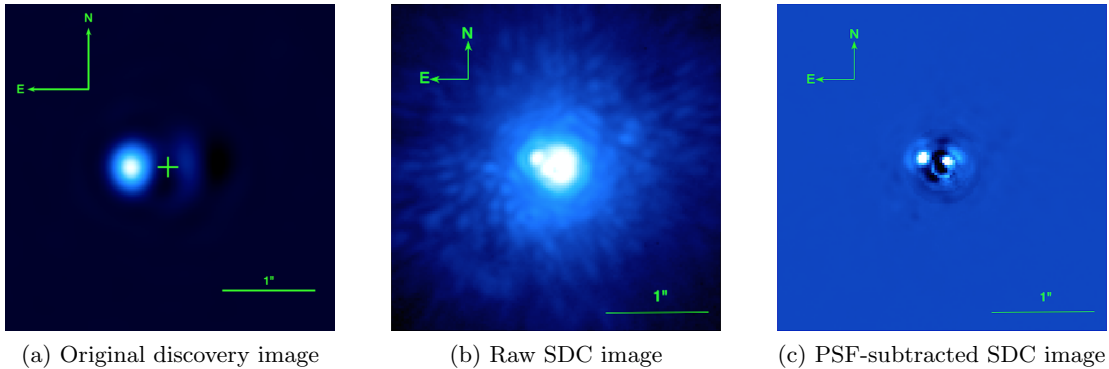

(a) Original discovery image     (b) Raw SDC image     (c) PSF-subtracted SDC image

Fig. 11.—: Epsilon Cephei b. (a) The original discovery image, from Mawet et al 2011 (Mawet et al. 2011a), using a 1.5 m well-corrected subaperture of the Hale telescope. (b) Raw (no reference subtraction) SDC image, dual vortex mode, 15s of 10 median combined frames. (c) Classic PSF subtraction of (b). In the SDC images, the first Airy ring is visible around the companion.

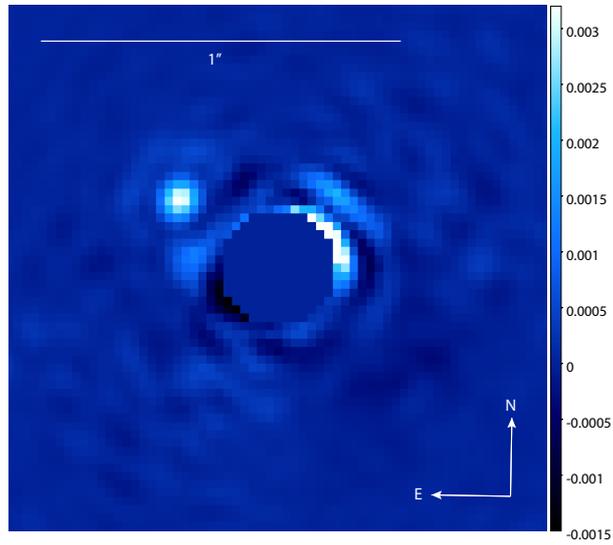

Fig. 12.—: The PSF-subtracted coronagraphic image of delta Andromeda b, dual vortex mode. This image first appeared in Ref. 1.

– 24 –